\documentclass[aps,prl,showpacs,twocolumn,superscriptaddress]{revtex4}
\usepackage[english]{babel}
\usepackage[T2A]{fontenc}
\usepackage[koi8-r]{inputenc}
\usepackage{amssymb,amsmath,amsfonts,amsthm}
\usepackage{mathrsfs}
\usepackage[mathcal]{eucal}

\usepackage[dvips]{graphicx}
\usepackage[colorlinks]{hyperref}

\input{epsf}

\begin{document}

\title{Phase diagram of the $t-t'$ Hubbard model taking into account spin-spiral waves and phase separation at finite temperatures}

\author
{A.K.~Arzhnikov}
\email{arzhnikof@bk.ru}
\affiliation
{Physical-Technical Institute, Ural Division of RAS, Izhevsk, Russia}
\author
{A.G.~Groshev$_{}^{}$}
\email{groshev_a.g@mail.ru}
\affiliation
{Physical-Technical Institute, Ural Division of RAS, Izhevsk, Russia}

\date{\today}

\begin{abstract}

The effect of temperature on the magnetic phase separation and the parameters of spin-spiral waves (SSW) is studied 
using a two-dimensional (2D) single-band $t-t'$ Hubbard model and the Hubbard-Stratonovich transformation. Both commensurate 
(antiferromagnetic (AF)) and incommensurate (helical) magnetic phases are considered. It is shown that the temperature 
significantly affects the collinear and helical magnetic phases. With an increase in the temperature, the 
phase-separation (PS) regions (AF$+[Q,Q])$, $([Q,Q]+[Q,\pi])$ get substantially reduced but new regions 
$([Q_{1},\pi]+[Q_{2},\pi])$, (AF$+[Q,\pi])$ arise. The results are used for the interpretation of the magnetic properties 
of cuprates.

\end{abstract}


\maketitle


The discovery, in the late 1980s, of high-temperature superconductivity giving rise to new extensive studies of the phenomenon 
of superconductivity, has also regenerated a vivid interest in quantum low-dimensional magnetism. At present most researchers are 
inclined to believe that magnetic interactions are responsible in great part (if not completely) for the high-temperature 
superconductivity. Apparently, the high-temperature pseudogap observed in superconductors is also connected with magnetism, rather 
than with the Cooper pairs \cite{Rui-Hua_2011}. Theoretical description of magnetism in these systems is still far from complete. 
As is shown experimentally \cite{Rui-Hua_He_2011}, the itinerant magnetism should enter as an essential component into theoretical 
models. Of great importance therewith is the two-dimensionality \cite{Shirane_1987, Kastner_1998}. The authors of paper 
\cite{Igoshev_2010} have studied the ground-state phase diagram of the 2D single-band Hubbard model. The calculations were performed 
in the framework of the mean field approximation with consideration of both commensurate (ferromagnetic and AF) and incommensurate 
(SSW) magnetic states. The results obtained made it possible to explain some peculiarities of the behavior of superconducting cuprates, 
and clarified the role of spatial magnetic phase separation (PS). Further investigation of the model at finite temperatures 
\cite{Arzhnikov_2011} has shown the necessity of taking into account the next-nearest-neighbor hopping, and revealed a significant 
influence of thermodynamic fluctuations on the boundaries of PS regions. 

In this paper we consider the 2D single-band Hubbard model at finite temperatures with allowance for the next-nearest-neighbor hopping. 
Note a controversial point that arises in studying the 2D models. It is related to the well known statement that magnetic order is 
lacking in 2D systems at finite temperatures (except for the Ising-type models). As a rule, the researchers which use the 2D Hubbard 
model for describing cuprates assume that the weak magnetic coupling between the Cu layers suppresses anomalous growth of transverse 
magnetic fluctuations and stabilizes the order, ensuring a quasi-two-dimensional behavior of the magnetic characteristics. However there 
may be an alternative mechanism of suppressing the anomalous growth of magnetic fluctuations through the charge degree of freedom and 
the volume Coulomb interaction. In this case one can speak of a 2D magnetic subsystem. The problem of the existence of 2D magnetism is 
similar to that of the 2D crystals in which the anomalous growth of lattice fluctuations is suppressed owing to strong anharmonicity 
of the longitudinal and transverse lattice vibrations \cite{Fasolino_2007}. In view of the above, in this work we use a special 
approximation which makes it possible to prevent anomalous growth of magnetic fluctuations and stabilize the magnetic order of the 2D 
model, leaving the question of the existence of 2D ordered structures unsettled. It should be noted that the results of such 
approximations are useful even in the absence of long-range magnetic order, as they may be used for the description of the dominant 
thermodynamical magnetic fluctuations at fixed temperatures. Another controversial point of our study is the choice of possible 
incommensurate magnetic structures. We have restricted our consideration to the helical magnetic phases (SSW), leaving aside the 
collinear spin-density waves (SDW). Usually, the findings obtained in neutron experiments give no possibility to distinguish between 
SSW and SDW (see, e.g., \cite{Yamada_1998, Matsuda_2000, Fujita_2002, Matsuda_2002}, except in a few cases such as the spin-polarized 
experiments on Fe-As monocrystals \cite{Rodriguez_2011} where the existence of SSW has been established unambiguously. Ab initio 
calculations show that SSW may be realized in many systems on basis of the transition metals \cite{Lizarraga_2004}. A direct comparison 
of the SSW and SDW energies within the 2D Hubbard model shows that in some cases the SDW are energetically preferable 
\cite{Tmirgazin_2012}. At the same time the obtained charge redistribution over the sites in SDW is too large, and the Coulomb interaction, 
not accounted for in the model, should increase the SDW energy, making these states unfavorable. 

\section{Model}

We adopt the Hamiltonian of the Hubbard model on the squared lattice
${\hat H}={\hat H}_{0}+{\hat H}_{int},$
\begin{equation}
\label{eq:hamiltonian}
{\hat H}_{0}=\sum_{ijs}t_{ij}\hat{c}_{is}^{+}\hat{c}_{js}^{},\,\quad
{\hat H}_{int}=U\sum_{j}\hat{n}_{j\uparrow}\hat{n}_{j\downarrow},\,
\end{equation}
where  $t_{ij}=-t$ for the nearest-neighbor sites $i,j$, $t_{ij}=t'$ for the next-nearest neighbors, 
$\hat{n}_{js}=\hat{c}_{js}^{+}\hat{c}_{js}^{}$ is the electrons number operator with spin projection 
$s=\uparrow,\downarrow $ on site $j$, $\hat{c}_{js}^{+} (\hat{c}_{js}^{})$ is a creation (annihilation) 
electronic operator with spin projection $s$ on site $j$. $U$ denotes the interaatomic Coulomb 
interaction on site.

We consider the helical-type magnetic structures corresponding to the magnetization vector uniformaly 
rotating in the polarization plane when moving from one site to another. This magnetic structure is 
characterized by the magnitude and direction of the wave vector $Q=[Q_{x}, Q_{y}]$. Generally, the 
wave vector $Q$ does not coincide with a reciprocal lattice vector and turns out to be incommensurate 
\cite{Igoshev_2010,Arzhnikov_2011}. The superposition of the helical wave and the ferromagnetic component 
perpendicular to the polarization plane has a large energy \cite{Timirgazin_2009} and it is not considering 
here. 

The thermal properties of the system are determined by the partition function of the grand canonical 
ensemble
\begin{eqnarray}
\label{eq:Z}
Z=Tr\left[T_{\tau}\exp\left\{-\int_{0}^{\beta}{\hat H}_{}(\tau)d\tau\right\}\right].\,
\end{eqnarray}
Here the symbol $Tr$ denotes summation over the comlete set of quantum states, $T_{\tau}$ denotes the 
time ordering operator, $\beta =1/k_{B}T$ denotes the inverse temperature, ${\hat H}_{}(\tau)$ is the 
${\hat H}-\mu{\hat n}$ operator in the interaction representation where $\mu$ and ${\hat n}$ are the 
chemical potential and the total electron number operator, respectively. 

Partition function (\ref{eq:Z}) for the many-particle problem is transformed into that of the single-particle 
problem with time-dependent fictitious fields, by means of the functional integral method through the use 
Hubbard-Stratonovich transformation \cite{Izymov_1987}. To reproduce the results of the generalized Hartree-Fock 
approximation at the ground state \cite{Igoshev_2010,Arzhnikov_2011} we need to use the two-field functional 
integral method in the static approximation \cite{Hassing_1973} introducing the spin and charge auxiliary 
fictitious fields, ${\bf v}_{j}$ and $\zeta_{j}$. In addition, we adopted the saddle point approximation for the 
charge auxiliary fictitious field $\zeta_{j}$ what corresponds to the replacement of the $\zeta_{j}$ by its value 
$\zeta_{j}^{0}({\bf v}_{j})$ at the saddle point. Using $\zeta_{j}^{0}({\bf v}_{j})$, we minimize the thermodynamic 
potential at the fixed configuration ${\bf v}_{j}$, but neglect the charge fluctuations. In this work, we take into 
account only the longitudinal spin fluctuations specifying the direction of the spin auxiliary fictitious field 
${\bf v}_{j}$ parallel to the magnetization vector ${\bf m}_{j}$ at each site.

The partition function is expressed by the mean field approximation for the single-particle Green function 
${\hat G}^{}(z)=[z-{\hat {\cal H}}_{MF}]^{-1}$ \cite{Wang_1969,Gorkov_1969}, where ${\hat H}_{MF}$ is the mean field 
approximation for the Hubbard Hamiltonian (\ref{eq:hamiltonian}). Finally, we introduce the self-energy of electron 
in the effective medium $\Sigma_{} $, which is determined by the average with respect to the amplitude of spin fictitious 
field $v$ Green function ${\tilde G}_{}=<[1-G_{}^{}V_{}(v)]^{-1}G_{}^{}>$. The effective self-energy $\Sigma $ and 
${\tilde G}$ are found using the self-consistent matrix equation in the single-site coherent potential approximation
\begin{equation}
\label{eq:CPA}
{\tilde G}_{}(z)=<\left[1-G_{}^{}V_{}(v)\right]^{-1}G_{}^{}>=G_{}^{}(z-\Sigma ).
\end{equation} 
After introducing these apprpximations, the partition function is represented in the form of an integral over the 
amplitude of spin fictitious field $v$
\begin{equation}
\label{eq:Omega}
Z=\exp\left\{-\beta(\Omega[\Sigma]+\Delta\Omega) 
\right\},\,
\end{equation}
\vspace {-5mm}
\begin{eqnarray}
\Omega[\Sigma]=\Omega_{MF}-
\frac{1}{\beta}\sum_{n}ln{\,det [1+\Sigma_{}(i\omega_{n}){\tilde G}_{}(i\omega_{n})]},\quad
\nonumber\\
\Delta\Omega=-\frac{1}{\beta}ln{\int d v\exp{\left\{-\beta\Delta\Omega(v)\right\}}},\qquad\qquad\quad
\nonumber\\
\Delta\Omega (v)=
\frac{1}{\beta}\sum_{n}ln{\,det [1-{\tilde G}_{}(i\omega_{n})(V_{}(v)-\Sigma_{}(i\omega_{n}))]}
\nonumber\\
+
U\left[v^{2}+\zeta_{}^{2}(v)\right].\qquad\qquad\qquad\qquad\qquad\quad
\nonumber
\end{eqnarray}
Here $\Omega[\Sigma]$ is the thermodynamic potential for the effective medium, which is determined by the effective 
self-energy $\Sigma$, and $\Delta\Omega$ is the fluctuating part of the thermodynamic potential, 
$\Omega_{MF}=-1/\beta\ln{Tr\exp{[-\beta({\hat H}_{MF}-\mu{\hat n})]}}$ is the mean field approximation for the 
thermodinamical potential of electrons, $\omega_{n}=\pi(2n+1)/\beta $ are the Matsubara frequencies for the Fermi 
particles. For brevity, site subscript in (\ref{eq:Omega}) is omitted. 

We mentioned earlier that the ground state in this approximation is the Hartree-Fock state, which is formally 
validated in the limit of the small parameter $U/t$. In our theoretical approach, the analysis of the approximations 
in terms of the small parameter at nonzero temperature is an extremely complicated problem, which is still unsolved.  
Nevertheless, the static approximation is justified since we are not studying the superconducting properties of our 
systems and the strong correlations in the chosen range of parametrs at the electron density close to unity are not 
very significant. This confirmed by comparison of our results \cite{Igoshev_2010} with those obtaned by the dynamical 
claster methods \cite{Maier_2005}. 

In spite of the aforementioned essential approximations, further analysis of expression (\ref{eq:Omega}) is 
possible only by using numerical methods. The self-consistent solution to (\ref{eq:CPA}) and (\ref{eq:Omega}) allows 
us to calculate all magnetic properties of our system under the condition that the magnetic state with the minimum 
thermodynamic potential is chosen.

\section{Results}

All calculations were carried out at $U/t = 4.8$, $t'/t=0.2$. For these values various magnetic phases are realized 
\cite{Igoshev_2010}, and the characteristics of the electron states of superconducting cuprates $(La_{(2-x)}Sr_{x} CuO_{4})$  
\cite{Lichtenstein_2000} may be described well enough. We restricted ourselves to the electron concentrations less than unity, 
because in a wide range of larger concentrations  the system exhibits a stable AF ordering \cite{Igoshev_2010}. 
Figure 1 presents the $T-n$ phase diagram (hereafter we shall use a dimensionless temperature $T\to k_{B}T/t)$. Thick 
solid lines correspond to the phase transitions between states with different magnetic order. The phase transition from 
the ordered magnetic state to a paramagnet (P) is a second-order transition, all the other are first-order transitions. 
This is well illustrated by the dependence of chemical potential on the electron concentration at $T=0.02$ Fig.~2 which 
exhibits a inflection when passing from AF to the diagonal $[Q,Q]$ phase, and from $[Q,Q]$ to the parallel $[Q,\pi]$ phase. 
Instability of the chemical potential (negative derivative with respect to concentration) results in spatial magnetic 
phase separation (PS) whose boundaries are determined by the Maxwell rule:
\begin{equation}
\label{eq:MAXWELL}
\int_{n1}^{n2}\left[\mu(n_{1})-\mu(n)\right] dn=0,\,
\end{equation}
where ${n1}$ and ${n2}$ are the boundaries of the PS region. A comparison of the PS region boundaries at zero and finite 
temperatures shows that with increasing temperature the PS regions narrow down, being replaced by homogeneous states. 
Besides there arise three new PS regions: $([Q_{1},\pi] + [Q_{2},\pi])$, $([Q,Q] + [Q,\pi])$, and (AF$\,+\,[Q,\pi])$ which were 
lacking at zero temperature. The characteristics of the magnetic phases and their partial ratios inside the PS regions are 
completely determined by their values at the boundaries of these regions. Figures 3, 4, 5, 6 present the parameters of 
spiral magnetic structures at the PS boundaries (AF$\,+\,[Q,Q])$, $([Q,Q] + [Q,\pi])$, $([Q_{1},\pi]+[Q_{2},\pi])$, and 
(AF$\,+\,[Q,\pi])$. These plots may be used to determine the mean local magnetization at site $<m>$, the mean absolute 
magnitude of local magnetization $<|m|>$, the spiral wave vector Q, and the partial phase ratio for any point inside the 
PS region. 

Of interest is a rather strong $T$ dependence of $Q$ in the phase mixture (AF$\,+\,[Q,Q])$, see Fig.~3. In experiments 
on cuprates close to half filling a region, generally referred to as "spin glass" (SG), is observed \cite{ Kastner_1998}. 
The PS (AF$\,+\,[Q,Q])$ with a strong temperature dependence of vector $Q$ found in our study can be easily associated  
with the experimentally observed SG region. Note that the ratio between the temperatures $T_{N}$ and $T_{g}$ corresponding 
to the P$\,\to\,$AF and AF$\,\to\,$ SG transitions, respectively, for $n= 0.98 $ is approximately equal to $12$ 
\cite{Matsuda_2002}, which agrees with the ratio of temperatures $T_{N}$ and $T_{PS}$ for the P$\,\to\,$AF and 
AF$\,\to\,$(AF$\,+\,[Q,Q])$ transitions, respectively, in our calculation Fig.~1. Recall that the calculated absolute 
values of the transition temperatures are overestimated because of the model and computational approximations made 
(see Section 1). The distinctive feature of the PS regions $([Q_{1},\pi]+[Q_{2},\pi])$ Fig.~5 and (AF$\,+\,[Q,\pi])$ 
Fig.~6 consists in a large difference between the mean local magnetic moments in the phase mixture, the absolute magnitudes 
of local magnetic moments being approximately equal. At high temperatures this fact can be interpreted as the effect of 
long-lived fluctuations against a paramagnetic-state background \cite{Shirane_1987}. Besides, the superparamagnetic behavior 
experimentally observed in chemically homogeneous Fe-Al alloys at high temperatures \cite{Arzhnikov_2008} is easily explained 
when taking into account the possibility of spatial magnetic separation between SSW and the ferrimagnetic phase. Note that 
low-temperature  neutron experiments have revealed in these alloys the existence of SDW \cite{Noakes_2003} the parameters of 
which are fairly well described in ab initio calculations using SSW 
\cite{Bogner_1998, Arzhnikov_2008}. 

The temperature and concentrational behavior of the PS region $([Q,Q]+[Q,\pi])$ reproduces two significant experimental facts 
\cite{Fujita_2002}. A decrease in the number of electrons results in the following sequence of transitions: 
$[Q,Q]\,\to\,([Q,Q]+[Q,\pi])\,\to\,[Q,\pi]$, and an increase in the temperature gives rise to the transition 
$([Q,Q]+[Q,\pi])\,\to\,[Q,\pi]$, see Figs.~1, 4.  It should be noted that quantitative agreement with the experimental concentrations 
of the transitions cannot be expected in our calculations, as they are extremely sensitive to the parameters $U/t$ and $t'/t$. 
An insignificant decrease in $U/t$ from $4.8$ leads to a considerable shift of these transitions towards concentrations close 
to unity, and at $U/t<4$ the concentrational transition connected with the $[Q,Q]$ phase totally disappears, see \cite{Igoshev_2010}. 

The authors of paper \cite{Igoshev_2010} have pointed out the proportional dependence of the wave vector on the electron 
concentration, which is in agreement with the experiment \cite{Fujita_2002} and can be easily interpreted in terms of the spin-spiral 
structure formation. At finite temperatures this dependence is retained, see, e.g. Fig.~2. Moreover, in the considered range of 
concentrations n, a general tendency for an increase in the SSW wave vector with temperature is observed. This is true both for each 
of the PS phases Figs.~3-6 and for the homogeneous states Fig.~7. However note that in the PS region the mean wave vector can diminish 
with increasing temperature. This is the case, for example, in the PS $([Q_{1},\pi]+[Q_{2},\pi])$ because of a change in the partial 
phase ratio, namely, a decrease with increasing temperature in the portion of the phase with larger wave vector Fig.~5. Of particular 
interest is the temperature transition metal-isolator near the concentration equal to unity. 

Figures 8 and 9 show the density of states for the AF and $[Q,Q]$ phases in the PS (AF$\,+\,[Q,Q])$ region at temperatures $0.02$ 
and $0.002$. It is seen that for $T=0.02$ all the phases are metallic, and AF phase may be considered as a conductor, since it has a 
lower electron density at the Fermi level as compared to the $[Q,Q]$ phase Fig.~8(a). With decreasing temperature the AF phase actually 
becomes isolator, and the conductivity of the $[Q,Q]$ phase increases Fig.~9(a). The metal-isolator transition experienced by the 
AF phase is a Slater-type transition \cite{Imada}. The total density of states for temperatures $0.02$ and $0.002$, and concentration 
$n=0.96$ is shown in Fig.~8(b) and Fig.~9(b). In both cases the total density of electron states at the Fermi level is different from 
zero. It is obvious however that at low temperature the system is practically an isolator, because at this temperature and concentration 
the portion of the AF phase relative to the $[Q,Q]$ phase exceeds $1/2$ (the percolation threshold in the 2D case being equal to $1/2$). 
Thus our system will undergo a metal-isolator transition in the concentration range $1\div 0.95$, but this is a Slater-type transition 
of percolation character. We did not account for the strong electron correlations which could modify the boundaries and features of 
the region of the transition described. It is also evident that with a decrease in the electron concentration (hole doping) the role 
of correlations substantially grows. The Mott transitions induced by the strong correlations in optimally doped and overdoped cuprate 
regions have been considered in detail in the review \cite{Lee_2006}.
\begin{figure}[t!]
\begin{center}
\includegraphics[scale=0.56]{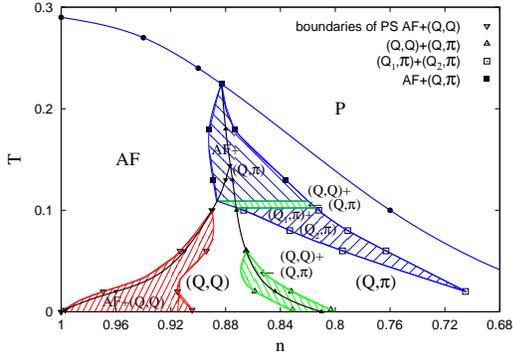}
\caption{(Color online) Magnetic phase diagram at $U/t=4.8$ and $t'/t=0.2$. 
Blue bold line denote the Neel temperature (second-order phase transition), black bold 
lines denote first-order phase transitions calculated without regard for PS, shaded areas 
denote PS regions.}
\end{center}
\end{figure}
\begin{figure}[t!]
\begin{center}
\includegraphics[scale=0.56]{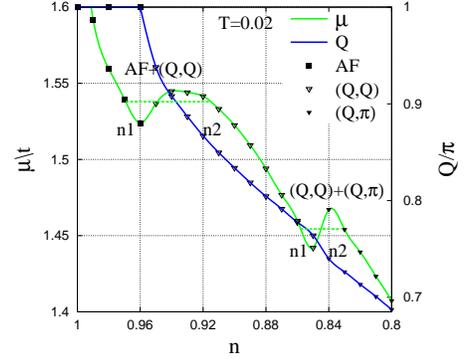}
\caption{(Color online) Chemical potential $\mu$ (left axis) and wave vector $Q$ (right 
axis) versus the electron density $n$ at $U/t=4.8$ and $t'/t=0.2$. Dashed lines denote $n$ dependence 
of the $\mu$ in the PS regions.}
\end{center}
\end{figure}
\begin{figure}[t!]
\begin{center}
\includegraphics[scale=0.56]{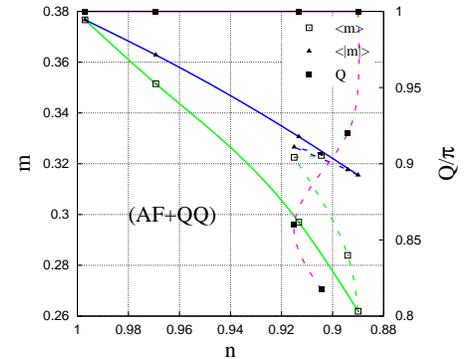}
\caption{(Color online) Electron number dependence of the average local magnetic moment $<m>$ 
and its absolute value $<\vert m\vert >$ (left axis), wave vector $Q$ (right axis) along the boundary 
of AF$\,+\,[Q,Q]$ PS region at $U/t=4.8$ and $t'/t=0.2$. Dashed lines - $[Q,Q]$ phase, bold lines - AF.}
\end{center}
\end{figure}
\begin{figure}[t!]
\begin{center}
\includegraphics[scale=0.56]{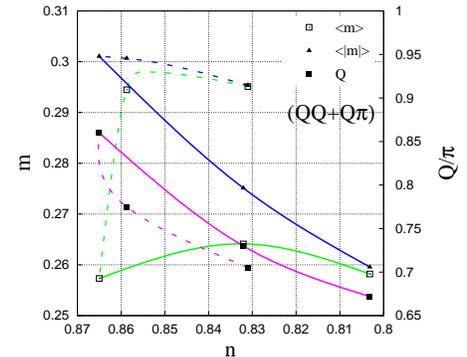}
\caption{(Color online)  $[Q,Q]+[Q,\pi ]$ PS region. Dashed lines - $[Q,Q]$ phase, bold lines - 
$[Q,\pi ]$. Notations are the same as in Fig.~3.}
\end{center}
\end{figure}
\begin{figure}[t!]
\begin{center}
\includegraphics[scale=0.56]{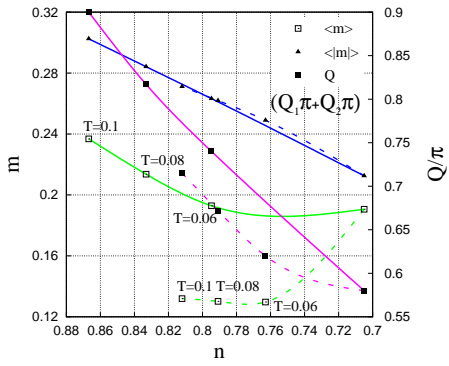}
\caption{(Color online) $[Q_{1},\pi]+[Q_{2},\pi]$  PS region. Dashed lines - top boundary whith 
$[Q,\pi]$ phase, bold lines - lower boundary whith $[Q,\pi]$ phase. Notations are the same as in Fig.~3.}
\end{center}
\end{figure}
\begin{figure}[t!]
\begin{center}
\includegraphics[scale=0.56]{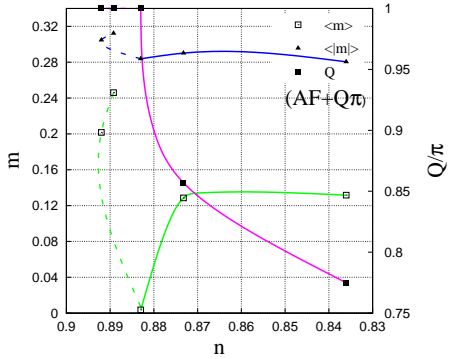}
\caption{(Color online)  AF$\,+\,[Q,\pi ]$ PS region. Dashed lines - AF, bold lines - $[Q,\pi ]$. 
Notations are the same as in Fig.~3.}
\end{center}
\end{figure}
\begin{figure}[t!]
\begin{center}
\includegraphics[scale=0.56]{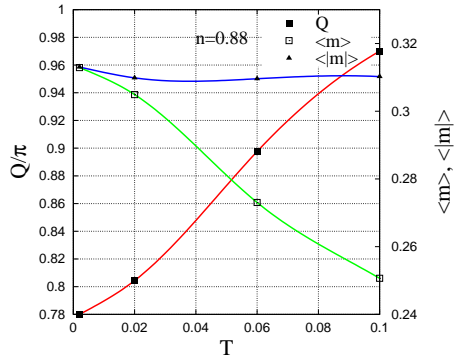}
\caption{(Color online) Temperature dependence of the wave vector $Q$ (left axis), average local magnetic 
moment $<m>$ and its magnitude $<\vert m\vert >$ (right axis) at $U/t=4.8$, $t'/t=0.2$ 
and $n=0.88$.}
\end{center}
\end{figure}
\begin{figure}[t!]
\begin{center}
\includegraphics[scale=0.56]{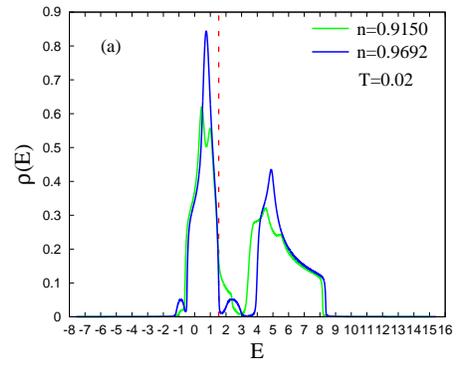}
\includegraphics[scale=0.56]{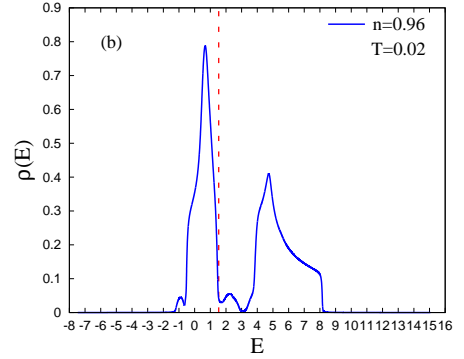}
\caption{(Color online) Electron density of states EDS in the AF$\,+\,[Q,Q]$ PS region. 
(a) Porcion EDS on the boundaries of PS at $T=0.02$. (b) Total EDS at $T=0.02$ and $n=0.96$. 
Dashed line - Fermy level $U/t=4.8$, $t'/t=0.2$}
\end{center}
\end{figure}
\begin{figure}[t!]
\begin{center}
\includegraphics[scale=0.56]{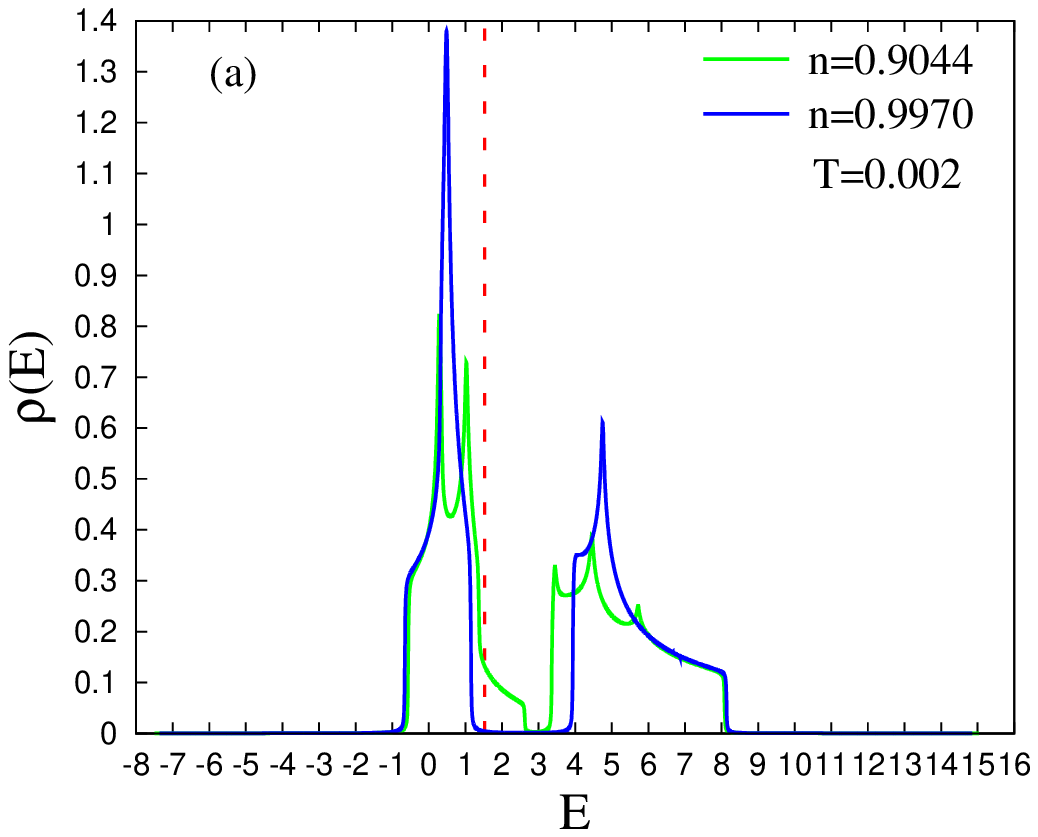}
\includegraphics[scale=0.56]{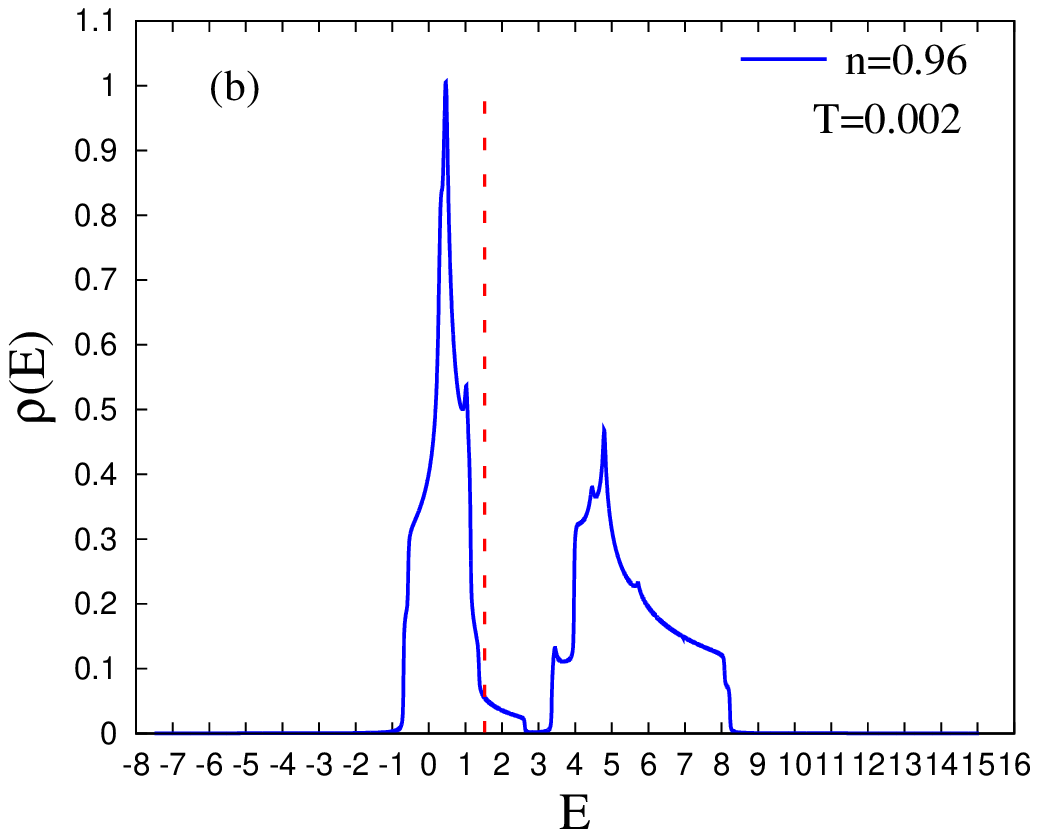}
\caption{(Color online) (a) $T=0.002$, $n=0.9044$ and $n=0.9970$. (b) $T=0.002$ and $n=0.96$. 
Notations are the same as in Fig.~8.}
\end{center}
\end{figure}

\section{Conclusion}

Thus, based on 2D Hubbard model, we have studied the formation of helical waves (SSW) and the magnetic PS at finite temperatures. 
The calculated $T-n$ phase diagram Fig.~1 shows that the PS regions existing at $T=0$ with increasing temperature get much narrower, 
and there arise new PS areas with different symmetry of the wave vectors SSW. The behavior of the wave vectors, mean local magnetic 
moments, and mean absolute magnitudes of the local moments is described for both the homogeneous SSW states and the phases forming 
the PS region. It is shown that the mean local magnetic moments of different phases inside the PS region may considerably differ in 
magnitude and temperature behavior. The wave vectors for all SSW increase with temperature. The proportionality of the SSW wave 
vectors to the number of electrons (holes) is retained in a wide range of temperatures. 

Taking account of the model and mathematical approximations made, the results obtained may be used to describe quasi-two-dimensional 
systems or the parameters of thermal magnetic fluctuations in the range of electron concentrations and temperatures where the effect 
of the strong correlations and system dynamics is not too significant. 

A comparison of the results obtained with the available experimental data for high-temperature compounds $La_{(2-x)}Sr_{x}$ $CuO_{4}$ 
shows a good semiquantitative agreement as to the behavior of the magnetic characteristics in the half-filling region with small hole 
concentration. The ratio between the temperatures $T_{N}$ (P$\,\to\,$AF transition) and $T_{g}$ (AF$\,\to\,$ "spin glass"\, transition) 
at $n=0.98$ approximates  to 12 \cite{Matsuda_2002}, which agrees with the ratio of temperatures $T_{N}$ and $T_{PS}$ for the P$\,\to\,$AF 
and  AF$\,\to\,$(AF$\,+\,[Q,Q])$ transition, respectively, obtained in our calculation. The presence of the PS region $([Q,Q]+[Q,\pi])$, 
the sequence of transitions $[Q,Q]\,\to\,([Q,Q]+[Q,\pi])\,\to\,[Q,\pi]$ occurring with increasing concentration, and 
$([Q,Q]+[Q,\pi])\,\to\,[Q,\pi]$ taking place with increasing temperature coincide with the PS region $([Q,Q]+[Q,\pi])$ and the transition 
sequences observed experimentally \cite{Fujita_2002}. 

Taking into account the possibility of a great difference in the mean local magnetic moment between the phases forming the PS region, one 
can explain the superparamagnetic behavior of chemically homogeneous alloys Fe-Al \cite{Arzhnikov_2008}.

Obviously, allowance for the strong correlations and the system dynamics will affect the quantitative characteristics of the PS regions 
and the SSW parameters. However, as demonstrated by a comparison of $T_{N}$ obtained in our calculations and in those performed 
within the dynamic cluster approximation (DCA) \cite{Maier_2005}, the difference proves to be not very large.

\acknowledgments

This work was supported by the Russian Foundation for Basic Research (project No.~12-02-00632) and Ur Br RAS (project No.~12-U-2-1021).

\end{document}